\documentstyle[prl,aps,twocolumn]{revtex}
\input{epsf}
\begin{document}

\title{Fermi Surface as the Driving Mechanism for Helical 
Antiferromagnetic Ordering in Gd-Y Alloys} 

\author{H.M. Fretwell,$^{1,2}$ S.B. Dugdale,$^{1,3}$ M.A. Alam,$^{1}$
D.C.R. Hedley,$^{1}$ A. Rodriguez-Gonzalez,$^{1}$ and S.B. Palmer$^{4}$}

\address{$^{1}$H.H. Wills Physics Laboratory, University of Bristol, Tyndall 
Avenue, Bristol BS8 1TL, United Kingdom}

\address{$^{2}$Department of Physics, University of Illinois at
Chicago, Chicago, IL 60607, USA}

\address{$^{3}$D\'epartement de Physique de la Mati\`ere Condens\'ee,
Universit\'e de Gen\`eve, 24 quai Ernest Ansermet, CH-1211 Gen\`eve 4,
Switzerland} 
\address{$^{4}$Department of Physics, University of Warwick, Coventry 
CV4 7AL, United Kingdom}

\date{\today}
\maketitle
\begin{abstract}
The first direct experimental evidence for the Fermi surface (FS) driving
the helical antiferromagnetic ordering in a gadolinium-yttrium alloy is
reported. The presence of a FS sheet capable of nesting is revealed, and
the nesting vector associated with the sheet is found to be in excellent
agreement with the periodicity of the helical ordering.
\end{abstract}

\pacs{71.18.+y, 78.70.Bj, 71.20.Eh, 87.59.Fm}

A considerable body of theoretical and indirect experimental evidence
indicates that the geometry of the Fermi surface (FS) drives a
variety of ordering phenomena ; these include exotic magnetic ordering in
the rare earths and their alloys \cite{kasuya:66}, compositional ordering
concentration density waves in binary alloys \cite{gyorffy:83}, and
magneto-oscillatory coupling in magnetic multilayers separated by
non-magnetic spacer layers \cite{parkin:91}. Current theoretical
understanding suggests that the ordering is governed by the ``nesting'' of
specific sheets of FS in the disordered state. Nesting 
describes the coincidence of two approximately parallel FS sheets when
translated by some distance in ${\bf k}$-space (i.e. by the ``nesting
vector'', $\bf Q$).  In the presence of nesting, the disordered phase
becomes unstable to an ordering modulation whose period is inversely
proportional to the relevant nesting vector. However, in many cases the
relevant features of the FS have never been directly observed.

Despite intense current interest in the concept of FS nesting as the
driving force for modulating magnetic structures, there is a distinct dearth
of direct experimental information about FS topologies in the
relevant materials. This scarcity has mainly been due to the lack of a
suitable technique. With recent developments in positron annihilation
Fermiology (see e.g. \cite{west:95}), such studies are now possible. In
this Letter, we provide the first direct evidence that a Gd-Y alloy which
orders antiferromagnetically does indeed contain a FS sheet that has
nesting properties. We compare the nesting vector directly calipered from
the FS topology with those inferred via neutron diffraction on specimens
from the same batch.  We also provide preliminary discussion of our
results in terms of the temperature dependence of the nesting
vector in the helical phase and effects of the transitions on this nested
FS.

A classic example of such a FS-driven ordering is the helical
antiferromagnetic ordering in many of the heavier rare earths (e.g. Tb, Dy,
Ho, Er). Here, the magnetic moments align in the basal planes with a
rotation of the moment vectors in successive planes with a
periodicity that is predicted to arise from the FS topology.  The ordering
has its origin in the coupling of the localised 4$f$ moments via the
Ruderman-Kittel-Kasuya-Yosida (RKKY) indirect exchange interaction
involving the conduction electrons \cite{kasuya:66}. The connection between
the conduction electrons' ability to establish magnetic order and the FS of
the disordered paramagnetic state is most easily understood in terms of the
wave-vector ($\bf q$) dependent susceptibility,
\begin{eqnarray}
\chi(\bbox{q}) \propto
\sum_{\bbox{k},j,j'} \frac{\vert M(\bbox{k},\bbox{k+q}) \vert^{2}
f_{kj}(1-f_{k+q+Gj'})}
{\epsilon_{j'}(\bbox{k}+\bbox{q}+\bbox{G})-
\epsilon_{j}(\bbox{k})}.
\label{suscept}
\end{eqnarray}
Here $M$ are the matrix elements involving the conduction and $f$ electron
wave functions, $f_{kj}$ are the Fermi--Dirac distribution functions for
reduced wave-vector ${\bf k}$ and band $j$, $\epsilon_{j}(\bbox{k})$ are
the single particle energies and ${\bf G}$ is a reciprocal lattice
vector. Subject to other constraints, the maximum in $\chi(\bbox{q})$
determines the most stable magnetic structure. If the maximum in
$\chi(\bbox{q})$ occurs at ${\bf q}=0$, the material orders
ferromagnetically. If the maximum occurs at a non-zero
${\bf q} = {\bf Q}$, then a more complex arrangement of the magnetic moments,
such as helical antiferromagnetic order, takes place. The latter may
happen if there are large parallel sections of FS to guarantee a
sufficient number of terms in the sum with vanishingly small denominators
at the nesting vector ${\bf Q}$. A so-called ``webbing'' feature
\cite{loucks:66,mattocks:78,dugdale:97} in the FS of most of the
rare earths provides the required parallel surfaces for nesting which drives
the magnetic ordering.

Helical antiferromagnetic ordering is also observed in Gd-Y alloys in a
certain composition and temperature range. The Gd FS does not
contain the webbing feature \cite{west:98} and Gd orders
ferromagnetically below 293K. The transition metal Y, on the other hand,
possesses a strong webbing sheet \cite{loucks:66,mattocks:78,dugdale:97}
but owing to the lack of magnetic moments is a paramagnet at all
temperatures. Addition of small amounts (of the order of 0.5 at.\%) of
Tb \cite{child:68,wakabayashi:74} or Er
\cite{caudron:90} leads to the appearance of long range magnetic structures
with ${\bf Q}$ vectors close to the nesting vector in the webbing feature
in Y. The helical phase observed in a range of Gd-rich Gd-Y alloys is
assumed to arise from a combination of magnetic moments contributed by Gd
and the Y-induced nesting character of the FS. Gd-Y provides an
ideal alloy system to study the generic magnetic behaviour in the rare
earths because of the availability of good quality single crystal samples
of sufficient size. Further, the concentration versus temperature magnetic
phase diagram in these alloys is well established and is rich in
interesting features, many of which are common to other rare earths
\cite{bates:85,melville:92}. In
the concentration range of 30---40 at.\% Y, the alloy shows a helical phase
where the helix periodicity, often quoted as the interlayer turn angle of
the basal plane moment vector, shows a reversible decrease with
temperature. This may imply a temperature-dependent nesting vector
(i.e. $T$-dependent changes in the webbing FS sheet). In the
concentration-temperature regime which supports the antiferromagnetism,
the application of a modest magnetic field (a few times 10$^{-2}$ tesla)
along the $c$-axis leads to a ferromagnetic alignment of the moments along
this axis. The specimen reverts to its antiferromagnetic state once the
magnetic field is switched off. In addition to the suggestion that the Fermi
surface of the magnetically disordered paramagnetic phase drives the
helical ordering, there is the issue of the effect of helical ordering on
the electron energy bands and therefore the impact on the FS
itself.

The measurements of the FS topology were conducted via the
so-called 2-Dimensional Angular Correlation of electron-positron
Annihilation Radiation (2D-ACAR). A 2D-ACAR measurement yields a 2D
projection (integration over one dimension) of the underlying two-photon
momentum density, $\rho(\bbox{p})$. Within the independent particle model,
\begin{eqnarray}
\rho(\bbox{p}) &=& \sum_{\bbox{k},j} \vert \int 
d\bbox{r}\psi_{k,j}(\bbox{r})\psi_{+}(\bbox{r}) \exp (-
i\bbox{p}.\bbox{r})\vert ^{2} \nonumber\\
&=& \sum_{j,\bbox{k},\bbox{G}} 
n^{j}(\bbox{k}) \vert C_{\bbox{G},j}(\bbox{k}) \vert ^{2} 
\delta(\bbox{p}-\bbox{k}-\bbox{G}),
\end{eqnarray}
where $\psi_{k,j}(\bbox{r})$ and $\psi_{+}(\bbox{r})$ are the electron and
positron wave functions, respectively, and $n^{j}(\bbox{k})$ is the electron
occupation density in ${\bf k}$-space in the $j^{\mbox{th}}$ band. The
$C_{\bbox{G},j}(\bbox{k})$ are the Fourier coefficients of the
electron-positron wave function product and the delta function expresses
the conservation of crystal momentum. $\rho(\bbox{p})$ contains information
about the occupied electron states and their momentum $\bbox{p} = \hbar
(\bbox{k +G})$ and the FS is reflected in the discontinuity in
this occupancy at the Fermi momentum $\bbox{p_{F}} = \hbar (\bbox{k_{F} +
G})$. As already pointed out, the 2D-ACAR spectra represent projections of
$\rho(\bbox{p})$, and the full 3D density can be reconstructed using
tomographic techniques from a small number of projections with integrations
along different crystallographic directions
\cite{cormack:63,cormack:64,sznajd:90}. Finally, if the effects of the
positron wave function (Eq. 2) are small such that the
$C_{\bbox{G},j}(\bbox{k})$ are almost independent of ${\bf k}$, the full 3D
${\bf k}$-space occupation density can be obtained by folding back the 3D
$\rho(\bbox{p})$ into the first Brillouin zone (BZ) according to the
so-called LCW prescription \cite{lock:73}.
By these means one is able to directly `image' the FS in its 
full 3D form.

The sample under investigation was a Gd$_{0.62}$Y$_{0.38}$ single crystal which
undergoes a helical antiferromagnetic transition below $\sim$200K where the
nature of the helical phase and its periodicity has been extensively
studied by one of us \cite{melville:92}. As part of a comprehensive
programme, we have also studied the FS topology of the two pure elements Gd
\cite{west:98,dugdale:96} and Y \cite{dugdale:97,dugdale:96}. In each case,
the 3D $\rho(\bbox{p})$ and subsequently the 3D ${\bf k}$-space occupancy
were reconstructed from five projections measured at 7.5 degree intervals
encompassing the 30 degrees between the directions \mbox{$\Gamma$-M} and
\mbox{$\Gamma$-K} (in the irreducible wedge of the hexagonal BZ). Following
the usual processing of the measured 2D-ACAR spectra \cite{west:95}, they
were deconvoluted using a `Maximum Entropy'-based (MaxEnt) procedure
\cite{dugdale:94} to suppress the unwanted smearing due to experimental
resolution. The 3D $\rho(\bbox{p})$ was reconstructed from the measured
projections (both raw and deconvoluted, henceforth referred to as `raw' and
`MaxEnt') using Cormack's method \cite{cormack:63,cormack:64,sznajd:90}
before finally being folded back into the first BZ. The reconstruction
method exploits the crystallographic symmetry which allows the
reconstruction from only a few projections and has been rigorously tested
\cite{sznajd:90,dugdale:96} to show that no artefacts are introduced into
the data. A more detailed description of the procedures used here can be
found in \cite{sznajd:90} and further references therein.
Using a threshold criterion to differentiate
between the empty and occupied states \cite{manuel:82,dugdale:97} it is
possible to image the FS alone.

\begin{figure}
\epsffile{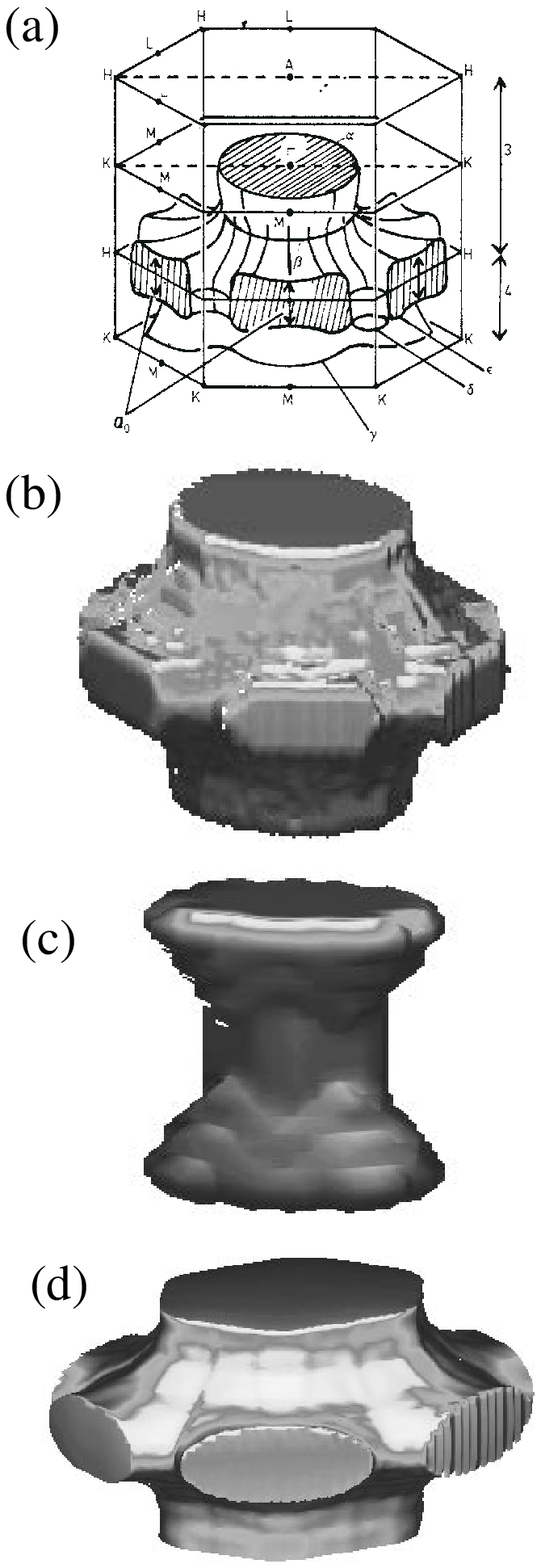}
\caption{(a) Calculated FS of Y \protect\cite{loucks:66}. (b) Measured FS
of Y ; note the presence of the `webbing' feature. (c) Measured FS of Gd ;
note that there is no webbing. (d) Measured FS of Gd$_{0.62}$Y$_{0.38}$,
showing a webbing feature.}
\label{FS3d}
\end{figure}

In Figure \ref{FS3d}, we show FS images of: (a) the calculated FS for Y
\cite{loucks:66}, where the webbing feature and the associated nesting
vector, ${\bf Q_{0}}$, is marked by the double arrow; (b) the measured
Fermi surface of Y \cite{dugdale:97} clearly showing the webbing feature;
(c) the measured FS of Gd \cite{west:98,dugdale:96} with the distinct lack
of the webbing feature as expected from its ferromagnetic ordering (d) the
current measurement of the alloy Gd$_{0.62}$Y$_{0.38}$ (in the disordered
paramagnetic phase) where the nesting feature is clearly visible. The
presented experimental FS images are extracted from MaxEnt deconvoluted
data. It is noteworthy that in the paramagnetic phase of the Gd-rich alloy,
the webbing feature is remarkably similar to that observed in pure Y. The
fact that the FS of pure Gd does not show the webbing feature while that of
the alloy has a strong nesting character clearly indicates that its helical
ordering is driven by the nesting of this sheet.

\begin{figure}
\epsffile{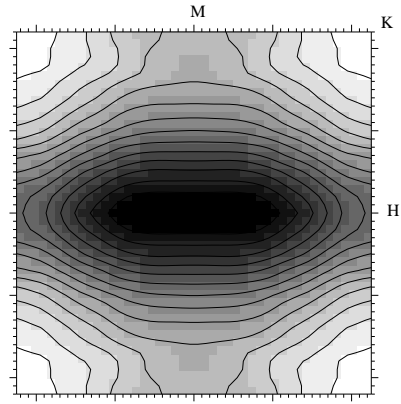}
\caption{Electron density in \mbox{H-L-M-K} plane. Black signifies holes,
and white represents electrons.}
\label{hlmk_raw}
\end{figure}

\begin{figure}
\epsffile{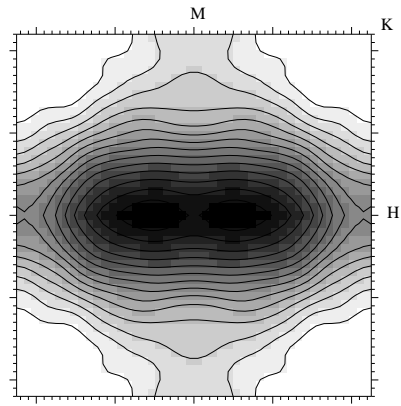}
\caption{Difference between MaxEnt \protect\cite{dugdale:94} and raw
electron densities in \mbox{H-L-M-K} plane.}
\label{hlmk_diff}
\end{figure}

We now investigate the webbing in greater detail. Figure \ref{hlmk_raw}
shows the section through the webbing in the alloy sample
(Fig. \ref{FS3d}d) in the \mbox{H-L-M-K} plane (on the face of the BZ in
Fig. 1a). Here, lows (holes) are shown as black and highs
(electrons) as white and the webbing is represented by the central region
of holes. If the raw and Maxent reconstructions are normalised to contain
the same number of electrons within the BZ and the raw reconstruction is
subtracted from the Maxent reconstruction, the distribution shown in Figure
\ref{hlmk_diff} results.  This procedure amounts to an edge-enhancement
that highlights the Fermi edges \cite{dugdale:94,dugdale:96}. It enhances
the discontinuity at the FS because it is in the vicinity of
these discontinuities that the resolution function has its greatest
smearing effect. Thus, the difference spectrum amplifies the signature of
the Fermi edge.

\begin{figure}
\epsffile{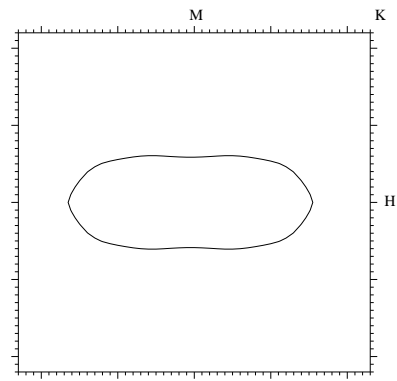}
\caption{The `zero contour' of Fig.\protect\ref{hlmk_diff}. Dugdale {\it et
al.} showed that this indicates the Fermi surface.}
\label{hlmk_zero}
\end{figure}

Dugdale {\it et al.} \cite{dugdale:94} proposed that the locus of points
where such a difference spectrum passes through zero defines the Fermi
surface, providing an accurate method for calipering it. The zero crossing
contour of Figure \ref{hlmk_diff} is shown in Figure \ref{hlmk_zero} which
again clearly shows the yttrium-like webbing feature and its nesting
properties. We estimate the width of the webbing parallel to the $c$-axis
as $0.53 \pm 0.02 \times \left( \pi \over c \right)$, a value remarkably
close to the nesting feature in pure Y of $0.55 \pm 0.02 \times \left( \pi
\over c \right)$ \cite{dugdale:97}. This would give rise to a period of
helicity which corresponds to an inter-plane turn angle of $47.7 \pm 1.6$
degrees between the orientations of the magnetic moments in successive
basal planes. As mentioned earlier, Bates {\it et al.}  \cite{bates:85}
measured the turn angle in the helical phase in a sample derived from the
same ingot via neutron diffraction by inspecting the distance between the
magnetic satellite Bragg peaks on either side of the nuclear peak along the
$c$-axis. These results showed that the measured turn angles were strongly
$T$-dependent, increasing linearly with temperature. The turn angles cannot
be measured in the paramagnetic phase through neutron diffraction owing to
the absence of the magnetic ordering and therefore the lack of the
satellite peaks. However, a linear interpolation of the $T$-dependence of
the turn angle returns a value of 48 degrees at 295K (the temperature of
our measurement) which is in excellent agreement with the value obtained in
our experiment.

Finally, it is worth mentioning our recent preliminary results from the same
alloy sample measured at $T=140$K, well within the thermodynamic
helical phase. In our experimental set-up, we use a magnetic field of
$\sim$0.8T to focus the positrons on to the specimen.  The field is applied
parallel to the $c$-direction of the sample. It was not possible to carry
out the measurements without the magnetic field owing to the small
dimensions of the sample, as a substantial fraction of the defocused
positrons annihilated in the sample holder and associated goniometer. As
noted above, the applied field would force the magnetic moments to align
along the easy axis giving rise to a $c$-axis ferromagnetic
state. Our preliminary analysis of the FS of the alloy sample in
this ferromagnetic state reconstructed from 3 projections again shows
(figure not shown owing to lack of space) distinct indication of the webbing
feature being present. 

If the periodic antiferromagnetic ordering introduced superzone boundaries
in the lattice, these may have distorted the FS feature in the webbing
sheet. However, in our low temperature data, such degeneracy would have
been lifted by the forced ferromagnetic ordering. If this is the case and
the relevant energy bands are not significantly affected, then one would
expect the webbing feature in the FS to be retained. Although theoretical
calculations are necessary to confirm this, it is a reasonable assumption.

In conclusion, we have shown for the first time the exisitence of a
nesting FS in a Gd-Y alloy ; this confirms theoretical predictions that
the FS topology in the paramagnetic state is responsible for the
antiferromagnetic ordering of the alloy.

The authors would like the thank the EPSRC (UK) for financial support. One
of us (SBD) is grateful to the Royal Society (UK) and the Swiss National
Science Foundation for the provision of a Research Fellowship.

\end{document}